\documentclass[%
aip,
jcp,%
amsmath,amssymb,%
preprint,%
]{revtex4-1}

\usepackage{graphicx}
\usepackage{verbatim}
\usepackage{amstext}
\usepackage{amssymb}
\usepackage{graphicx}
\usepackage{esint}
\usepackage{amsmath}

\begin{document}

\title{Spectrins in Axonal Cytoskeletons: Dynamics Revealed by Extensions and Fluctuations}
\author{Lipeng Lai}
\affiliation{Department of Chemistry, Massachusetts Institute of Technology, Cambridge, MA 02139}
\affiliation{MIT-SUTD Collaboration, Massachusetts Institute of Technology, Cambridge, MA 02139}
\author{Jianshu Cao}
\email{jianshu@mit.edu}
\affiliation{Department of Chemistry, Massachusetts Institute of Technology, Cambridge, MA 02139}

\date{\today}

\begin{abstract}

The macroscopic properties, the properties of individual components and how those components interact with each other are three important aspects of a composited structure. An understanding of the interplay between them is essential in the study of complex systems. Using axonal cytoskeleton as an example system, here we perform a theoretical study of slender structures that can be coarse-grained as a simple smooth 3-dimensional curve. We first present a generic model for such systems based on the fundamental theorem of curves. We use this generic model to demonstrate the applicability of the well-known worm-like chain (WLC) model to the network level and investigate the situation when the system is stretched by strong forces (weakly bending limit). We specifically studied recent experimental observations that revealed the hitherto unknown periodic cytoskeleton structure of axons and measured the longitudinal fluctuations. Instead of focusing on single molecules, we apply analytical results from the WLC model to both single molecule and network levels and focus on the relations between extensions and fluctuations. We show how this approach introduces constraints to possible local dynamics of the spectrin tetramers in the axonal cytoskeleton and finally suggests simple but self-consistent dynamics of spectrins in which the spectrins in one spatial period of axons fluctuate in-sync. 

\end{abstract}

\maketitle

\section{Introduction}

\label{sec:introduction}

The properties of a complex structure largely depends on the properties of its individual components and how these components interact with each other cooperatively in the structure. An understanding of the interplay between the three aspects are essential in various fields, such as architectures and designs of new materials. In this paper, we focus on cytoskeletons, known as dynamic biopolymer networks enclosed within the cell's membranes. 

The cytoskeleton is made of different filaments with various lengths and stiffnesses. The functions of living cells largely depend on the cytoskeleton in many aspects, such as the cellular growth or locomotion. Intensive studies have been conducted for both \textit{in vivo} and \textit{in vitro} systems, with respect to the relationships between the macroscopic properties of the biopolymer network and the elastic properties and local structures of its components (e.g.,  Gardel \textit{et al.}\citep{gardel04} and Lieleg \textit{et al.}\citep{lieleg10}). Various experimental methods, such as the neutron scattering or the small angle X-ray scattering, can be applied to measure the local structures of the cytoskeletons, or \textit{in vitro} filamentous networks. Here we primarily focus on the situation where a direct measurement of the dynamics of a network at the single molecule level is not available experimentally. This may be due to the smallness or complexity of the structure, or the complicated interactions involved. We specifically study the axonal cytoskeleton and demonstrate that how the observed elastic properties of the network and the single spectrin, reveal the dynamics at the level of single molecules, in our example, the cooperative dynamics of spectrins in the axons.

\begin{figure}
\centerline{\includegraphics[width=12cm]{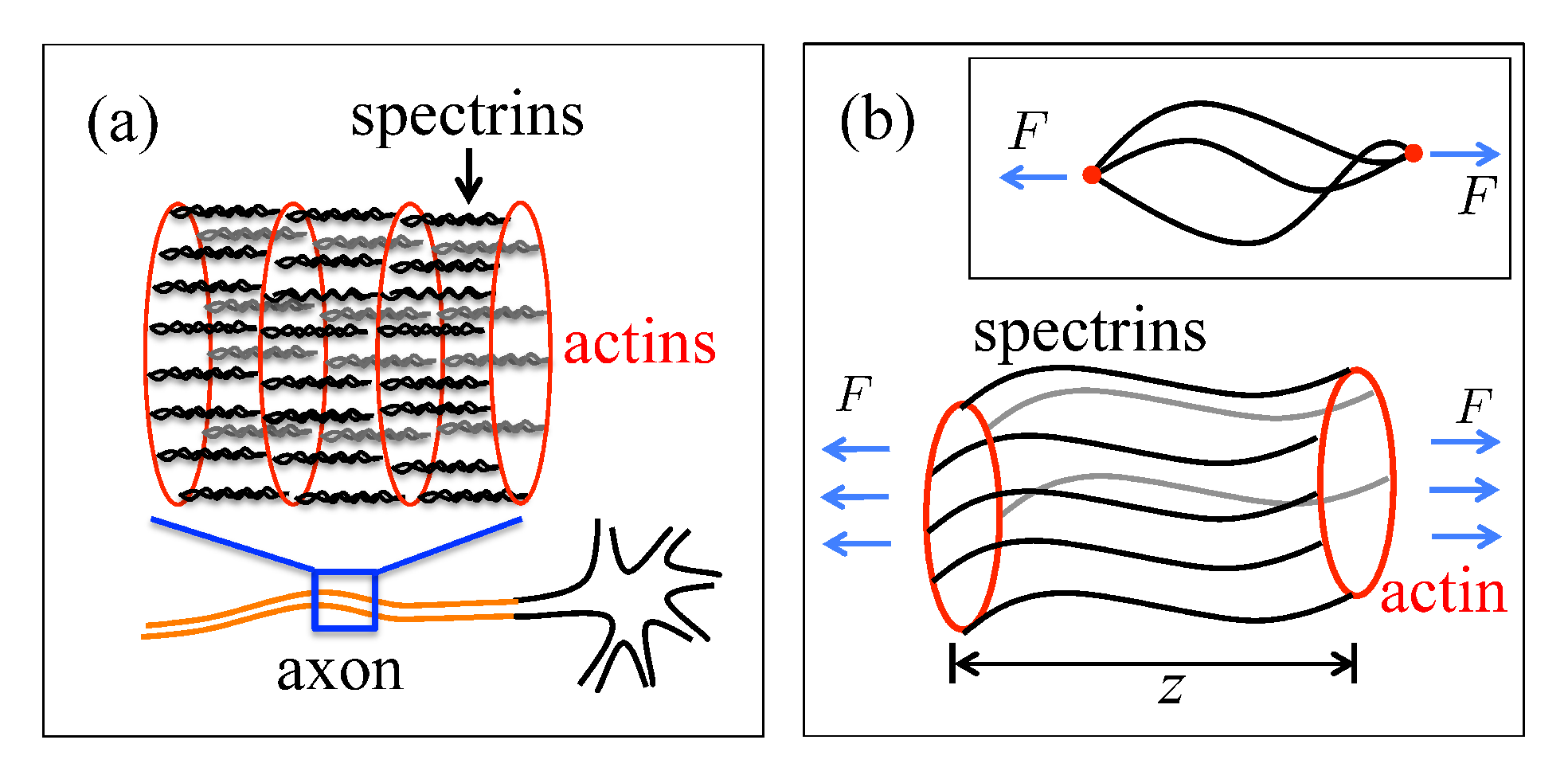}} \protect\caption{Sketch of the experimental observations and model assumptions. (a)
Sketch of the experimentally observed periodic structure in axonal
cytoskeleton\citep{xu2012}. Actin rings (red) are connected by spectrins
(black and gray) in parallel. (b) Sketch of the possible configurations
and dynamics of the spectrin tetramers in the axonal cytoskeleton.
Main figure: all the spectrins in the same period are considered to
bend in sync (see texts for details). Inset: illustration of the case
when spectrins fluctuate independently with the relative positions
of their ends on each side fixed.}
\label{fig:axon} 
\end{figure}

The axonal cytoskeleton in neuron cells has its own special structures and functions (e.g., Dennerll \textit{et al.}\citep{dennerll88}, Hammarlund \textit{et al.}\citep{hammarlund07}, and Galbraith \textit{et al.}\citep{galbraith93}). Recent experiments by Xu \textit{et al.}\citep{xu2012} revealed the novel 1-dimensional periodic cytoskeletal structure in axonal shafts, using the stochastic optical reconstruction microscopy (STORM). In such a structure, the axonal cytoskeleton resembles a hose tube periodically supported by rings that are formed by actins. Adjacent rings are connected by spectrin tetramers (Fig. \ref{fig:axon} (a)). The average and the fluctuation of the spacing between the rings are measured. We will show that, when a specific model is applied, these two quantities give us information about the dynamics of spectrins that connect two adjacent actin rings. 

The model we specifically apply here is the worm-like chain (WLC) model \citep{kratky49, marko95}. Providing an alternative view of the WLC model, we also introduce a more generic description of the energy form when the system can be coarse-grained as a linear curve, and show that this description reduces to the WLC model under certain conditions. This alternative view supports the extended applicability of the WLC model, from single molecule level to the network level, and also provides some insights into possible modifications to current models of slender systems. To study the axonal cytoskeleton, we apply the WLC model at both the single molecule and the network levels, and utilize the relations between average extensions and longitudinal fluctuations to limit the possible dynamics of spectrins. Our approach demonstrated in this paper will be useful when the observation of the dynamics in the network at the single molecule level is limited by experimental techniques. We hope  this approach can facilitate the study of other systems, when they are slender and can be coarse-grained as linear curves.

This paper is structured in the following way. In Section \ref{sec:model}, we will first introduce
the generic model when the composited network is coarse-grained
as a single smooth curve. We will show explicitly that this model reduces
to the WLC model under certain conditions. Then in Section \ref{sec:results},
analytical results from the WLC model are presented in the strong-stretching limit (one realization of the weakly-bending limit), focusing on the relations between the extensions and fluctuations projected to the force direction. The results will be applied to the axonal
cytoskeleton to demonstrate how these relations at the single spectrin and cytoskeletal network levels limit the possibilities of the cooperative dynamics of spectrins. Lastly, we
will briefly discuss several relevant topics and conclude our main
results.

\section{Model of a coarse-grained slender system and its energy}
\label{sec:model}

In this section, we first present a generic model for an arbitrary slender structure (a chain) in the weakly bending limit, being a single molecule or a network. The slender structure is coarse-grained as a smooth curve. This model can be reduced to the well-known worm-like chain (WLC) model under certain conditions. On the one hand, we use this model to provide an alternative understanding of the WLC model from a mathematical perspective via the fundamental theorem of curves\citep{erwin91, carmo76}. This understanding supports the applications of the WLC model and other relevant models at different scales, ranging from single molecules to any complex systems coarse-grained as linear curves. On the other hand, this model also offers insights into what modifications we can add to the current WLC model.

\begin{figure}
\centerline{\includegraphics[width=6cm]{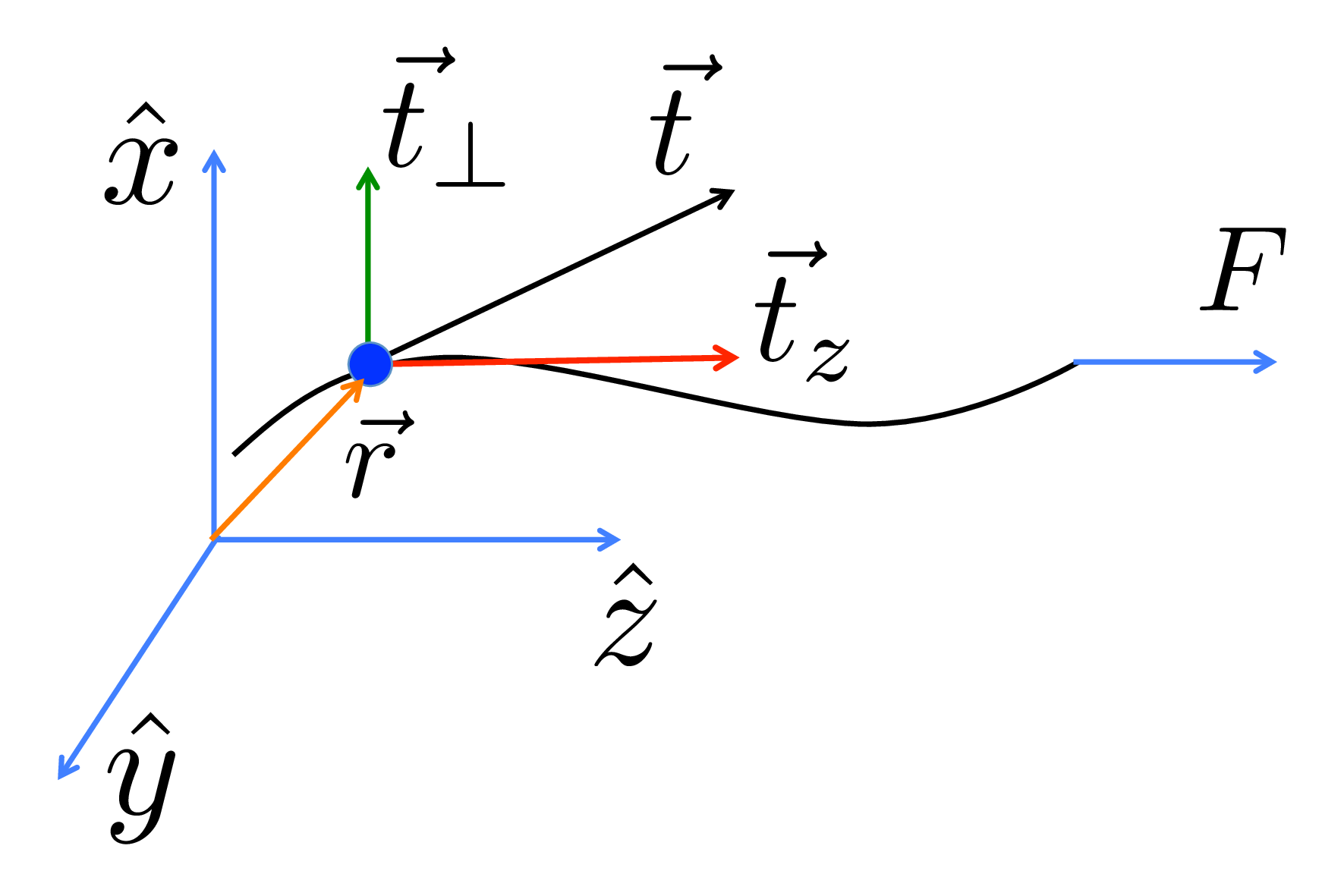}} \protect\caption{Sketch of a chain stretched by a single force $F$. Here a single
polymer or a polymer network that is thin and long is modeled as a
continuous 3-dimensional curve.}
\label{fig:setup} 
\end{figure}

Here we use a continuous description of a linear chain (stretched or unstretched, Fig. \ref{fig:setup}). The coarse-grained system is represented by a curve $\vec{r}(s)$, where $s$ marks the points on the curve and corresponds to the arc-length of the curve when it is unstretched. In a stretched state, the local strain along the curve is $u=d\tilde{s}/ds$ where $\tilde{s}$ is the arc-length of the stretched chain.

In general, the energy of the coarse-grained system is a functional of the curve and can be written as 
\begin{equation}
E=E_{e}[\{\vec{r}(s)\}]+E_{s}[\{\vec{r}(s)\}]
\end{equation}
where $E_{e}$ describes the portion of energy determined by the internal
coordinate $\text{\ensuremath{s}}$ (such as
the elastic energy) and $E_{s}$ depends on the global spatial coordinates
(such as the hydrodynamic interactions, the excluded volume effect, or external potentials).
$\{\vec{r}(s)\}$ denotes the set of the function $\vec{r}(s)$
and all orders of its derivatives with respect to $s$. Here we focus on $E_{e}$ first. Since $E_{e}$ only depends on the shape of the curve and the fundamental theorem of curves states that a smooth 3-dimensional curve is determined by its curvature and torsion, we can write $E_{e}$ as a functional of the curvature $\kappa(s)$, torsion $\tau(s)$, and the local stretching $\delta u(s) = u(s) - 1$. In the limit of weakly bending and small extension, we can expand the curve around the base state, the straight line, and formally write

\begin{equation}\label{eqn:energyexpand}
E_{e} = E_{e}^0+\intop_0^{L_0}\intop_0^{L_0} ds ds'\left(\frac{1}{2}V(s)^{T}H(u; s,s')V(s')\right)
\end{equation}

Here we restrict our model to only local interactions, and thus $V(s) = (\kappa(s), \tau(s), \delta u (s))$ and $H = \{h_{mn}(u, s)\delta(s-s')\}$ describing the interaction strengths. $L_0$ is the contour length of the unstretched system. It is interesting to notice how Eq. (\ref{eqn:energyexpand}) corresponds to existing models. The energy due to pure bending is $\int_0^{L_0} ds h_{11}\kappa(s)^2/2$ with $\kappa(s) = [(\partial^{2}\vec{r}/\partial s^{2})-(\partial^2\vec{r}/\partial s^2)\cdot(\partial\vec{r}/\partial s)(\partial \vec{r}/\partial s)/u^2]/u^2$ (e.g., Soda\citep{soda73}) and $h_{11}$ the bending rigidity. $\int_0^{L_0} ds h_{33}\delta u^{2}/2$ corresponds to the elastic energy due to stretching. The terms containing the torsion $\tau$ (e.g., $h_{22}\tau^{2}/2$)
reflect more complex features of the chain's local structures. It is straightforward to see that Eq.  (\ref{eqn:energyexpand}) reduces to the well-known worm-like chain model \citep{kratky49} when we only take into account the bending energy (i.e., $\tau(s) = 0$, $u(s) = 1$). The WLC model has been applied to different kinds of biopolymers like DNAs and microtubules, and has successfully explained various experimentally observed behaviors of long polymer chains, such as the force-extension relation (e.g., Marko \textit{et al.}\citep{marko95}) that is frequently measured in single-molecule experiments (e.g., Smith \textit{et al.}\citep{smith92}), the moments of the end-to-end distance distribution for a free chain (e.g., Schurr \textit{et al.}\citep{schurr00}), to name some of them. Modifications to the WLC model are also studied extensively. For example, different modifications have been proposed to explain the experimentally observed enhanced flexibility of DNA molecules at short length scales (e.g., Yan \textit{et al.}\citep{yan05} and Xu \textit{et al.}\citep{xuxinliang2013}).

Now we consider when an external force field that corresponds to a potential $\mathcal{V}$ is applied. Examples of such force fields include the case when the
system is stretched by a single force at one end with the other end
fixed or when it is immersed in a constant plug flow with one
end fixed. In the single force stretching case, the potential energy
can be written as $\mathcal{V}(z) = -Fz$, where $F$ is the magnitude of the force
and $z$ is the displacement of the free end relative to the fixed
end in the force direction. When we put the potential $\mathcal{V}(z)$ into the WLC model, the energy now reads:

\begin{equation}\label{eqn:wlc}
E=\intop_{0}^{L_{0}}\frac{h_{11}}{2}\kappa(s)^{2}ds-\intop_{0}^{L_{0}}Ft_{z}(s)ds,
\end{equation}
where $t_{z}$ is the component of the unit tangent vector of the
curve $\vec{r}(s)$ in the force direction. Previous studies of such a system primarily focused on the relations between the average end-to-end distance or extension $z$ as a function of the force applied\citep{marko95, hori07, purohit2008}. Some of them also showed the relations between the higher moments of $z$ as functions of the force \citep{su2010}. But there are instances, as we show later, in which the force is not measured experimentally. In this case, we can still use the relations between the average extension $z$ and its higher moments to obtain information from experimental data.


In the following section, we apply the WLC model with external force (Eq. (\ref{eqn:wlc})) to both single molecules and the cytoskeletal network. Then we combine the results obtained at different scales to infer the local dynamics of spectrins in the axonal cytoskeleton.

\section{Extensions and longitudinal fluctuations of a strongly stretched inextensible chain}
\label{sec:results}

\subsection{Analytical solutions}

The relation between the extension and variation of $z$ can be derived from the relations between the force and the moments of $z$. The latter is obtained by taking successive derivatives of the partition function $\mathcal{Z}$ of the model (Eq. (\ref{eqn:wlc})) with respect to the force $F$, where the partition function
\begin{eqnarray}
\mathcal{Z} & = & \int_{C}[\mathcal{D}\vec{t}]\exp[-\beta E]\nonumber \\
 & = & \int_{C}\left[D\vec{t}\right]\exp\left[{-\beta\int_{0}^{L_{0}}\left(\frac{h_{11}}{2}\left(\frac{d\vec{t}}{ds}\right)^{2}-Ft_{z}\right)ds}\right]\label{eq:partition}
\end{eqnarray}
is a weighted sum of all possible configurations. Here $\beta=1/k_{B}T$ with $k_B$ the Boltzmann constant and $T$ the temperature. In general, the partition function of a system coarse-grained as a linear curve corresponds to a path-integral\citep{vilgis2000}. Analytical results of $\mathcal{Z}$ are obtained in literature via various methods, such as mean field approaches\citep{ha97, winkler03}, or in the weakly bending limit\citep{yang03, hori07, purohit2008, su2010}. Here, related to the axon cytoskeleton, we are specifically interested in the limit when the force $F$ is strong, which is an example of the weakly bending limit. In this limit, the transverse
fluctuations (fluctuations perpendicular to the force direction) are
largely suppressed. The $z$-component of the tangent vector dominates
the other two components in $x$ and $y$ directions, i.e., $t_{x}\sim t_{y}\ll t_{z}$.
Because $|\vec{t}|=1$, we have $t_z\approx 1- (t_x^2+t_y^2)/2$. With this approximation, the partition function $\mathcal{Z}$ can be solved analytically by different techniques, such as by using the equal-partition theorem\citep{marko95, purohit2008, su2010}, or by using path-integral techniques\citep{yang03, hori07}. The final analytical expression of $\mathcal{Z}$ also depends on the boundary conditions we choose. In our case, since the spectrins form the wall of the tube of axons, without further experimental observations, it is natural to assume that the spectrins bind to actin rings with a right angle. Based on this symmetry argument, we have $t_x = t_y = 0$ at both ends ($s = 0$ and $L_0$) as our boundary conditions. It should be noted that, since different boundary conditions impose different constraints on the entropy, they may lead to measurable differences in the force-extension curve, especially when the contour length $L_0$ is comparable to the persistence length $l_p$. Here the persistence length $l_p = \beta h_{11}$ describes the competition between the bending energy and thermal energy and defines the correlation
length between two tangent vectors along the curve. The effects of different boundary conditions are discussed in details in previous studies\citep{hori07, purohit2008, su2010}. In our case, these effects should be negligible because the spectrins have a relatively large ratio of the contour length to the persistence length ($\sim 10$) and these effects are less significant when the force increases. However, more detailed experimental measurements may reveal such small differences and it will be interesting for future investigations.

With all the conditions specified, using path-integral techniques, we obtain the partition function
\begin{equation}
\mathcal{Z}=e^{\beta FL_{0}}\mathcal{Z}_{t}^{2}=e^{\beta FL_{0}}(h_{11}F)^{1/2}\left(\frac{\beta}{2\pi\sinh\sqrt{FL_{0}^{2}/h_{11}}}\right)\label{eq:partitionanalytic}
\end{equation}
Detailed calculations can be found in the appendix for completeness. the $n$th
moment of $z$ can be obtained by taking successive derivatives of
the partition function $\mathcal{Z}$ with respect to the force $F$:
\begin{equation}
\langle z^{n}\rangle=\frac{1}{\mathcal{Z}}\frac{\partial^{n}\mathcal{Z}}{\partial(\beta F)^{n}}.\label{eq:moments}
\end{equation}

Applying Eq. (\ref{eq:moments}) at $n=1$, we obtain the average
fractional extension of the chain:
\begin{equation}
\frac{\langle z\rangle}{L_{0}}=1-\frac{1}{2\beta}\left(\frac{1}{\sqrt{h_{11}F}}\coth\left(\sqrt{\frac{F}{h_{11}}}L_{0}\right)-\frac{1}{FL_{0}}\right).\label{eq:relativeextensioncomplete}
\end{equation}

\begin{figure}
\centerline{\includegraphics[width=12cm]{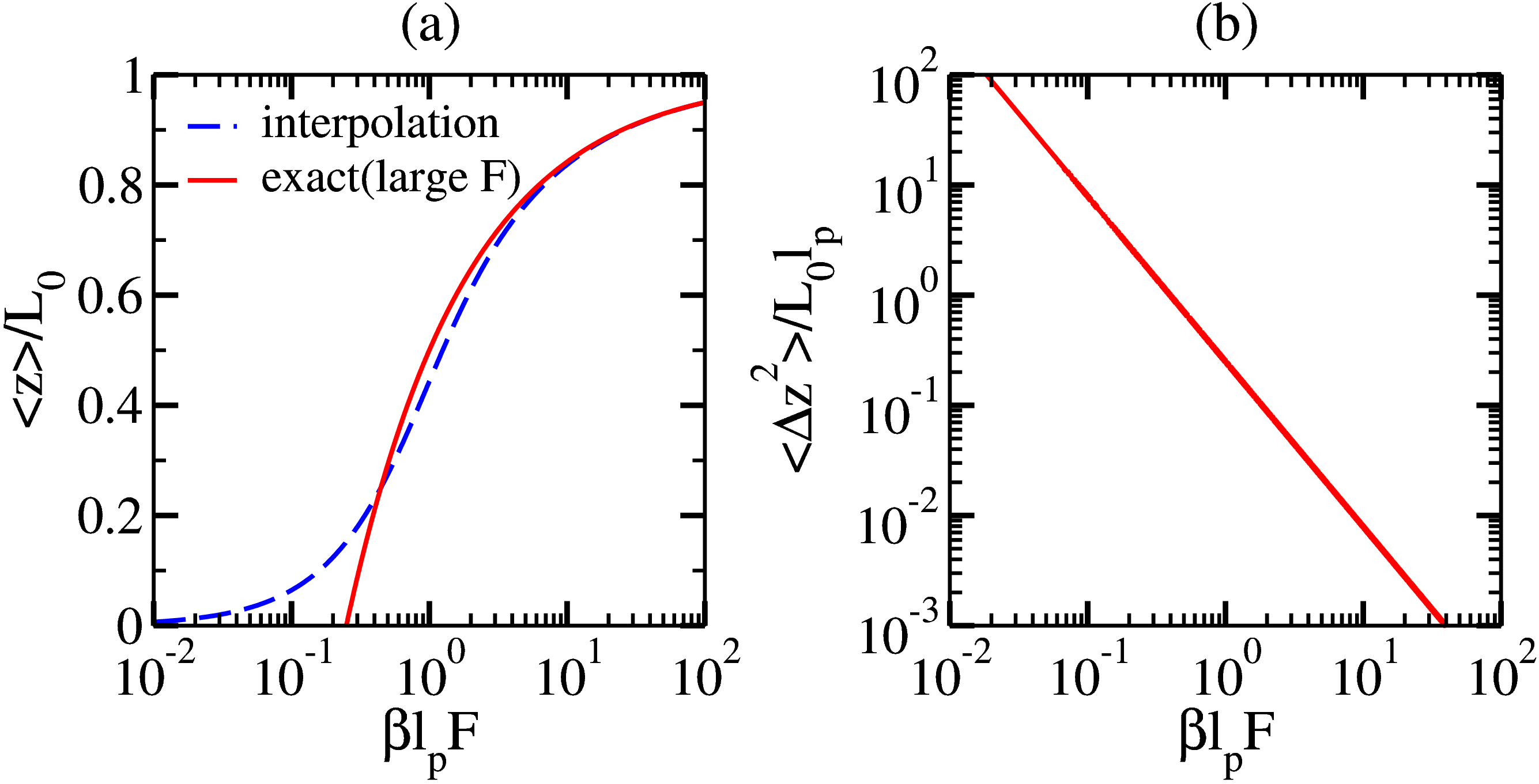}} \protect\caption{First two cumulants of the projected end-to-end distance as functions
of the external force $F$. (a) The relative average extension ($\langle z\rangle/L_{0}$)
as a function of the dimensionless force $\beta l_{p}F$ (red solid
curve). The blue dashed curve is the approximated interpolation formula
$\beta l_{p}F=\frac{\langle z\rangle}{L_{0}}+\frac{1}{4(1-\langle z\rangle/L_{0})^{2}}-\frac{1}{4}$
by Marko and Siggia\citep{marko95}. It shows that, as we expect,
these two results agree in the large force region ($\beta l_{p}F>1$).
(b) The dimensionless variance $\langle \Delta z^{2}\rangle/(L_{0}l_{p})$ as a
function of the dimensionless force $\beta l_{p}F$. In the large
force region, the variance decays as $F^{3/2}$.}

\label{fig:forcerelation-1} 
\end{figure}

In the strong-stretching (large force) limit, we only keep the leading
terms ($\sim1/\sqrt{F}$). So we have: 
\begin{equation}
\frac{\langle z\rangle}{L_{0}}=1-\frac{1}{2\beta}\left(\frac{1}{\sqrt{h_{11}F}}\right)=1-\frac{1}{2}\frac{1}{\sqrt{\beta l_{p}F}},\label{eqn:extension-1}
\end{equation}
which agrees with previous calculations\citep{marko95, hori07}. Fig. \ref{fig:forcerelation-1}
(a) shows the comparison between Eq. (\ref{eqn:extension-1}) and the
approximated interpolation formula from Marko and Siggia\citep{marko95},
and they agree with each other in the large force region $\beta l_{p}F>1$
as one should expect. This is the regime ($\beta l_{p}F>1$) where
Eq. (\ref{eq:relativeextensioncomplete}) and (\ref{eqn:extension-1})
hold because of the strong-stretching approximation used in calculating the partition function. 

Similarly, for the fluctuations (the second cumulant) $\langle\Delta z^{2}\rangle\equiv\langle z^{2}\rangle-\langle z\rangle^{2}$,
we have 
\begin{eqnarray}
\langle\Delta z^{2}\rangle & = & \frac{1}{\beta}\frac{\partial}{\partial F}\left(\frac{1}{\beta}\frac{\partial\ln\mathcal{Z}}{\partial F}\right)=\frac{1}{\beta}\frac{\partial\langle z\rangle}{\partial F}\nonumber \\
 & = & \frac{L_{0}l_{p}}{4\left(\beta l_{p}F\right)^{3/2}}\coth\left(\sqrt{\frac{F\beta}{l_{p}}}L_{0}\right)+O\left(\frac{1}{F^{2}}\right).
\end{eqnarray}
Again, in the strong-stretching limit, we have (Fig. \ref{fig:forcerelation-1}
(b)): 
\begin{equation}
\langle\Delta z^{2}\rangle=\frac{L_{0}l_{p}}{4\left(\beta l_{p}F\right)^{3/2}}.\label{eqn:variance-1}
\end{equation}
In principle, all the cumulants (or moments) of $z$ can be generated
from Eq. (\ref{eq:partitionanalytic}). 

\begin{figure}
\centerline{\includegraphics[width=12cm]{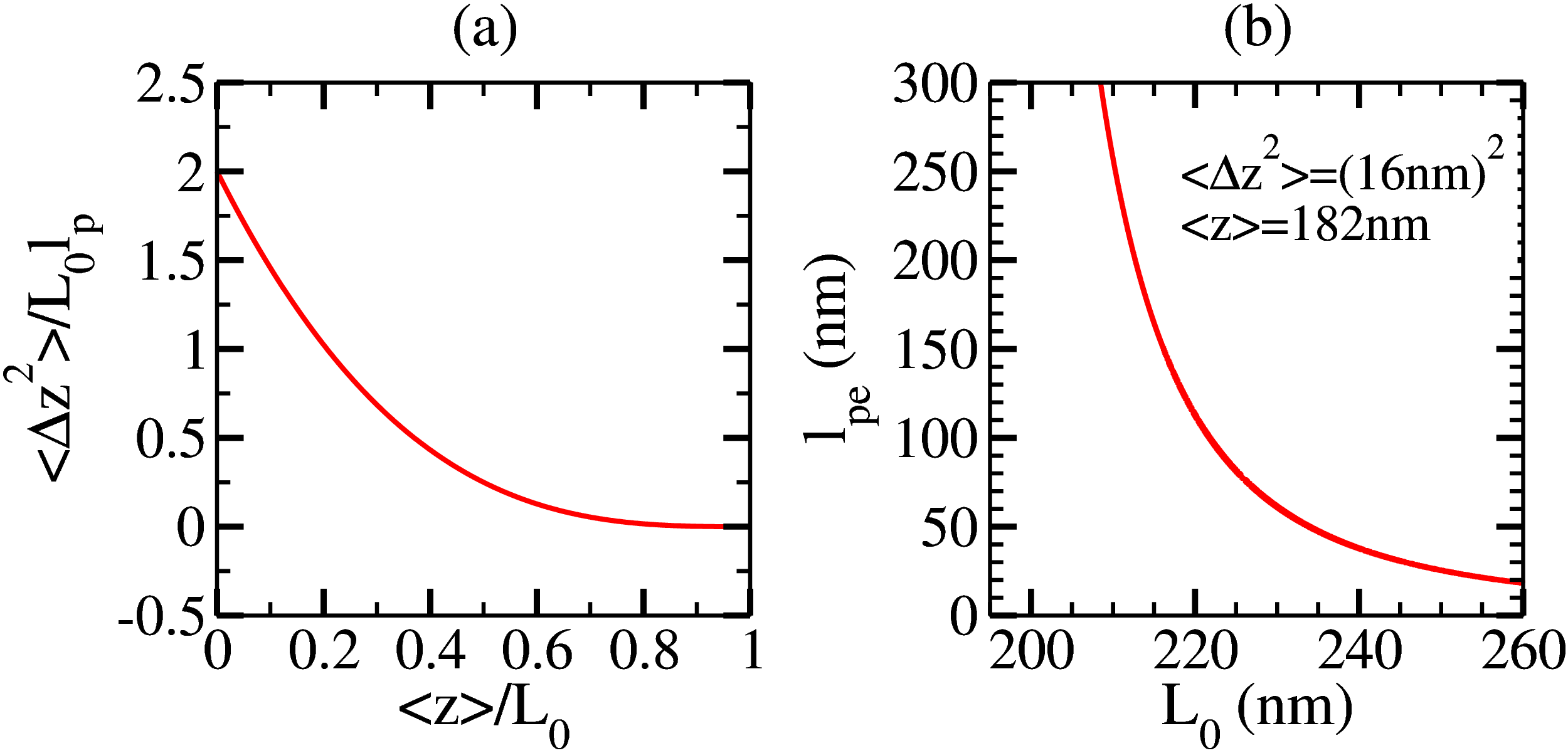}} \protect\caption{The relation between the first two cumulants and its constraint on
the contour length $L_{0}$ and the persistence length $l_{pe}$.
(a) The relation between the dimensionless variance $\langle \Delta z^{2}\rangle /L_{0}l_{p}$
and the relative average extension $\langle z\rangle/L_{0}$. (b) Given the average
extension $\langle z\rangle$ and variance $\langle \Delta z^{2}\rangle$ from experiments\citep{xu2012},
this curve shows the relation between the persistence length $l_{pe}$
and the contour length $L_{0}$. Here $l_{pe}$ is the effective persistence
length for a bundle of spectrins.}

\label{fig:experiment-1} 
\end{figure}

The force $F$ is not directly measured in the axon experiment\citep{xu2012}.
From Eq. (\ref{eqn:extension-1}) and Eq. (\ref{eqn:variance-1}), we
eliminate the dependence on the force $F$ and obtain that 
\begin{equation}
\langle\Delta z^{2}\rangle=2L_{0}l_{p}\left(1-\frac{\langle z\rangle}{L_{0}}\right)^{3}.\label{eqn:noforce-1}
\end{equation}
This relation includes two parameters: the contour length $L_{0}$
and the persistence length $l_{p}$, which are the material properties
of the spectrin. Eq. (\ref{eqn:noforce-1}) is consistent with the scaling
analysis discussed by Odijk when the longitudinal dispersion of DNA
in nano-channels is studied\citep{odijk09}. The scaling argument
establishes a sixth power law between the longitudinal variance $\langle\Delta z^{2}\rangle$
and the typical angle $\theta$ that the polymer makes with respect
to the $z$-axis, i.e., $\langle\Delta z^{2}\rangle\sim\theta^{6}$. In Eq. (\ref{eqn:noforce-1}),
if we notice that $\langle z\rangle/L_{0}\sim\cos\theta$ and use $\cos\theta\approx1-\theta^{2}/2$
in the strong-stretching or weakly bending limit, we get the same
sixth power law as the scaling argument. Fig. \ref{fig:experiment-1}
(a) shows the fluctuation $\langle \Delta z^{2}\rangle$ as a function of the average
extension $\langle z\rangle $.

\subsection{Application to experimental observations}

The relation between the extension and variation of $z$ (Eq. (\ref{eqn:noforce-1})) applies to both
single polymers and polymer networks when
the system can be coarse-grained as a linear smooth curve. Here we demonstrate
how we can apply the results to experimental observations and obtain
the information of local configurations of the axonal cytoskeleton. 

The example we are interested in here is the recent observation of
the 1-dimensional periodic structure in the axonal cytoskeleton \citep{xu2012}.
In such a structure, the axonal cytoskeleton resembles a hose tube
periodically supported by rings that are formed by actins. Adjacent
rings are connected by bundles of spectrin tetramers (Fig. \ref{fig:axon}
(a)). However, little is known about the dynamics of the spectrins
in such a network and how the interactions between spectrins and other components of
the cell (such as the membrane) affect the elastic properties of the cytoskeleton. Here
we focus on how the spectrins between two adjacent actin rings fluctuate relative
to each other. Two simple but different assumptions are considered. One assumption is that the spectrins in the same period fluctuate
in sync (Fig. \ref{fig:axon}(b)). Mathematically this means that
for any two spectrins between the same pair of actin rings, we have
$\vec{t}_{1}(s)=\vec{t}_{2}(s)$ and $\frac{d\vec{t}_{1}}{ds}(s)=\frac{d\vec{t}_{2}}{ds}(s)$
where the subscripts $1$ and $2$ differentiate the two spectrins
in the same period. The physical consequence of this assumption is
that now we can consider the spectrins in one spatial period as a
bundle, which behaves like a single chain with a new effective bending
rigidity $h_{11}'$. Since they bend in the same way, the bending
energy of the bundle will be $N$ times that of a single spectrin,
where $N$ is the total number of spectrins in the bundle. Thus we
have $h_{11}'=Nh_{11}^{s}$ and hence an effective persistence length
for the bundle $l_{pe}=Nl_{p}^{s}$, where $h_{11}^{s}$ and $l_{p}^{s}$
are the bending rigidity and persistence length of one single spectrin
tetramer respectively. The other assumption is that even though the
ends on each side of the spectrins are held relatively fixed to each
other, they can fluctuate independently in between (inset of Fig.
\ref{fig:axon}(b)). In this case, the entropy increases linearly
with the number of spectrins. Thus with an increasing number of spectrins,
the stronger entropic effect implies a shorter effective persistence
length. Later we will show that the experimental observations support
the first assumption. In both cases, we assume that the spectrins
have the same contour length $L_{0}$. To make a quantitative comparison
with the experiment, we further assume that the spectrins have a fixed
connection with the actins with a right angle, and the forces acting
on the spectrins can be effectively considered as from the actin rings
at the ends and the average direction of the forces are perpendicular
to the rings. We also assume that the planes of the actins are parallel
to each other, which implies that the spacing between adjacent rings
represents the end-to-end distance of the spectrins in the force direction.
The experimental measurements\citep{xu2012} suggest that the average
end-to-end distance projected to the force direction is about $182$
$nm$. This is close to the contour length of the spectrin tetramer
($\sim200$ $nm$) \citep{stokke85,svoboda92,li05}, and we expect
that our calculation in the strong-stretching limit should apply.
Some of the assumptions may not seem obvious without further experimental
data. Our point here is to demonstrate how the analytical results can be applied to both the single molecule and cytoskeleton levels to infer the dynamics of the components in the cytoskeletal network. The limitations of those assumptions will be discussed in the next section.


Using the experimentally measured average spacing $182$ $nm$ as
$\langle z\rangle$ and the variance of the spacing $(16$ $nm)^{2}$ as $\langle \Delta z^{2}\rangle$,
a relationship between the contour length $L_{0}$ and the effective
persistence length $l_{pe}$ is plotted in Fig. \ref{fig:experiment-1}
(b) using Eq. (\ref{eqn:noforce-1}). This curve represents a constraint
that is put on the possible values of $L_{0}$ and $l_{pe}$ by experimental
measurements. It shows that if we use a contour length of spectrins
around $200$ $nm$, then the experimental measurements predict an
effective persistence length of the axon cytoskeleton much larger
than the persistence length of single spectrin that is only around
$10$ -- $20$ $nm$. Thus experimental observations support the first assumption that the spectrins fluctuate in-sync in one spatial period. The reason of this dynamics is unknown
and probably has its root in the interaction between spectrins and
other components in the cell, such as the membrane or the microtubules.
To further check if this in-sync-fluctuation assumption is consistent
with experiments, we use the contour length $L_{0}$ and the persistence
length $l_{p}$ for single spectrins from literatures\citep{stokke85,svoboda92,li05} and
the measured average extension\citep{xu2012} $\langle z\rangle=182$ $nm$ to
predict the variance $\langle\Delta z^{2}\rangle$ using Eq. (\ref{eqn:noforce-1}) (Table
\ref{table1}). To get the effective persistence length $l_{pe}$,
we make a rather rough estimation of $N=12$ as an approximated total
number of spectrins between two adjacent actin rings. Experiments
suggest that $N$ is the order of $10$, and since we are only checking
the consistency here, slightly varying the value of $N$ will not
affect the results shown in Table \ref{table1} qualitatively.

\begin{table}
\begin{centering}
\protect\caption{\label{table1}The fluctuations predicted by Eq. (\ref{eqn:noforce-1}) using contour
lengths and persistence lengths from literatures\citep{stokke85,svoboda92,li05}.
We assume $N=12$ as the approximated total number of spectrins in
one spatial period and use $\langle z\rangle=182$ $nm$ for the average extension\citep{xu2012}.}

\par\end{centering}

\centering{} %
\begin{tabular}{ccccc}
\hline 
$L_{0}$ ($nm$)  & $l_{p}$ ($nm$)  & $l_{pe}$ ($nm$)  & $\sqrt{\langle \Delta z^{2}\rangle}$ ($nm$)  & Ref.\tabularnewline
\hline 
200  & 16.4  & 196.8  & 7.6  & Stokke \textit{et al.} \citep{stokke85} \tabularnewline
200  & 10  & 120  & 6  & Svoboda \textit{et al.} \citep{svoboda92}\tabularnewline
237.75  & 7.5  & 90  & 23.5  & Li \textit{et al.} \citep{li05}\tabularnewline
\hline 
\end{tabular}
\end{table}

The difference in the predicted fluctuations $\langle\Delta z^{2}\rangle$ in
Table \ref{table1} is probably due to different experiment or simulation
setups. But more importantly, Table \ref{table1} shows that the
experimentally observed variance ($16$ $nm$)$^{2}$ falls between
the fluctuations predicted by Eq. (\ref{eqn:noforce-1}), implying
the consistency between the experiments and our model assumptions. We can also reproduce the variance of $\langle\Delta z^{2}\rangle=$($16$ $nm$)$^{2}$ with parameters very similar to those in literatures (for example, using $L_{0}=214$ $nm$, $l_{p}=15$ $nm$, $N=12$, and $\langle z\rangle=182$ $nm$, we get $\langle \Delta z^{2}\rangle=(16$ $nm$)$^{2}$ according to Eq. (\ref{eqn:noforce-1})).

Thus, without introducing further complexity or more subtle mechanism,
we demonstrate that combining the relations between extensions and fluctuations at both the single molecule and cytoskeleton levels help us understand more about the dynamics of spectrins in the axonal cytoskeleton. Although we cannot rule out other possibilities, here we use the theoretical results to provide a possible simple explanation to the experimentally observed longitudinal fluctuations in the axonal cytoskeleton.

\section{Discussions}

Several topics deserve further discussions here. First of all, when
we apply our analytical results to the specific experiments of axonal
cytoskeletons, the details of the interactions between the cytoskeleton
and other components of the axon are ignored. However, the conclusion that the assumption of in-sync fluctuations of the spectrin tetramers, as one of possible configurations,
is self-consistent may imply that the averaged effect of those interactions
at leading order is to keep the spectrins fluctuates in sync. Here
we consider that the experimentally measured fluctuations is purely
caused by the thermal fluctuations of the spectrins. But in reality,
the fluctuations observed should include contributions from all aspects,
and only set an upper bound for the thermal fluctuations of spectrins.
In this paper, we used a largely simplified setup to demonstrate the
idea of how the macroscopic elastic properties of a coarse-grained
chain put constraints on its local dynamics and obtained a self-consistent
result. A more subtle modeling will definitely be useful with future
experimental support. It should be noted again that we cannot rule
out other possibilities solely using the theory, but the conclusion
here may provide some hints of what to look for in future investigations.

It should also be mentioned that the system we studied corresponds to an isotensional ensemble (or Gibbs ensemble). Its counter part is the isometric ensemble (or Helmholtz ensemble), in which the end-to-end distance is fixed a constant and the force fluctuates. The equivalence of these two ensembles are discussed in previous studies\citep{winkler10, manca2014}, especially in the thermodynamic limit. In a biological system, such as the axon cytoskeleton, one would expect both the end-to-end distances and the forces fluctuate. At current stage, the differences between the two ensembles for systems with finite length will not alter our conclusion here. But a realization of the isometric ensemble and more precise experimental measurements will determine the more appropriate boundary conditions {\textit{in vivo}}.

Here we studied the relations between the extensions and fluctuations when a slender object is stretched at the end. It will be intriguing to understand such relations in a more general setup. Examples include the case when the system is influenced by external flows or electric fields. Experimental, numerical, and analytical studies have been performed in this direction (e.g., Perkins \textit{et al.} \citep{perkins95}, Smith \textit{et al.}\citep{smith99}, Manca \textit{et al.}\citep{manca12}, and Yang \textit{et al.}\citep{yang03}). Stretching polymers using external fields is also an crucial technique in manipulating single molecules \citep{bustamante2000, hsieh11, wang12}. In future studies, we will focus on how our approaches presented here can help understand the dynamics of biopolymer networks influenced by external fields.

As a goal for longer terms, we would like to know how the local structures
and dynamics are related to the specific functions of certain biological
systems. For example, for the axonal cytoskeleton, similar structures
have been noticed in different situations, such as the skeleton of
snakes or the structure of hoses. As an analogy to those structures,
in neuron cells, the periodic actin rings may provide support against
the bending or transverse compression of the axons. But it deserves
further investigations to understand how the axon benefits from the
specific structure of the actin-spectrin network. If the actin-spectrin
network helps stabilize the membrane, then we can ask whether the
spacing between and the radius of the ring (or the ratio between these
two) observed experimentally correspond to an optimal solution of
a certain underlying mechanism, such as the buckling instability.
From a Physicist's perspective, if similar cytoskeletal structure
exists in other organisms, it will be more intriguing to see weather
the ratio between the ring's spacing and radius is universal across
different organisms. If the answer is yes, then it is natural to ask
what mechanism dictates this ratio.

Finally, although we specifically apply our results to the
axonal cytoskeleton, our derivations are based on a slightly more
general setup. We expect that our results may facilitate the study of other
biological or physical structures where the composited system can
be coarse-grained as a single smooth curve. On the other hand, in this paper
we treat the fluctuations as purely thermal. But it is known that,
in a living cell, the dynamical process is rather active with energy
input (e.g., from ATP) and dissipation (e.g., Kim \textit{et al.}\citep{kim13}).
So \textit{in vivo}, the configurational fluctuations may be affected
by the energy input rate, such as the ATP concentration. However,
the same approach applied in this paper may also be used in such an
active system. Given the observed fluctuations and the elastic properties
of the components, the details of the active process may be revealed
(e.g., Kim \textit{et al.}\citep{kim13}). 

\section{Conclusion}
To conclude, we investigated the interplay between the macroscopic properties of a composited system, the elasticity of its components and their local dynamics. We focus on slender systems that can be coarse-grained as linear curves. A generic model for such systems was introduced based on the fundamental theorem of curves. This model provides an alternative view of the well-known WLC model from a mathematical perspective and supports the applications of the WLC model at different scales, ranging from single molecules to complex networks. Then we applied the WLC model to both axonal cytoskeletons and individual spectrins, one component of the axonal cytoskeletons. Different from previous studies that primarily focus on the relations between forces and polymer conformations, we utilized the relations between the fluctuations and average extensions to understand the experiments where the force was not measured. Combining such relations on both the spectrin and the cytoskeleton levels suggest a simple but self-consistent local dynamics of spectrins, in which all the spectrins in one spatial period in the axonal cytoskeleton fluctuate in-sync. Although the validity of this dynamics requires further experimental and theoretical investigations, we used it as one example that shows how we can apply the WLC model or relevant models at different scales of slender systems and extract information from the comparison between theoretical predictions and experimental observations. We hope that the models and ideas presented here will help further studies of biological or physical polymer networks that can be coarse-gained as linear curves, and improve our understanding of the relations between the macroscopic and microscopic dynamics of complex systems.

\section*{Acknowledgements}

We thank Dr. Ke Xu for helpful discussions and inspiring comments,
and also thank Dr. Xinliang Xu for suggestions. We acknowledge the
financial assistance of Singapore--MIT Alliance for Research and Technology
(SMART), National Science Foundation (NSF CHE--1112825), and the Graduate
Fellows Program by Singapore University of Technology and Design and
MIT (to L.L).

\appendix

\setcounter{equation}{0}
\renewcommand{\theequation}{A\arabic{equation}}

\section*{APPENDIX: ANALYTICAL RESULTS OF THE PARTITION FUNCTION}

For completeness, here we outline the key steps in calculating the partition function $\mathcal{Z}$. More details can be found in literatures\citep{vilgis2000, yang03, hori07, purohit2008}. With the approximation $t_z\approx 1- (t_x^2+t_y^2)/2$, the fluctuations in transverse directions (i.e., $x$ and $y$) are decoupled and we can write the partition function (Eq. (\ref{eq:partition})) as 
\begin{equation}
\mathcal{Z}=e^{\beta FL_{0}}\mathcal{Z}_{x}\mathcal{Z}_{y}\label{eq:separatepartition}
\end{equation}

where $\mathcal{\mathcal{Z}}_{x}$ and $\mathcal{Z}_{y}$ have the
same form
\begin{equation}
\mathcal{Z}_{x}=\mathcal{Z}_{y}\equiv\mathcal{Z}_{t}=\int_{C}\left[Dt\right]\exp\left[{-\beta\left(\frac{h_{11}}{2}\int_{0}^{L_{0}}\left(\frac{dt}{ds}\right)^{2}+\frac{F}{2}t^{2}\right)ds}\right].\label{eq:zx}
\end{equation}

It is noted that $\mathcal{Z}_{x}$ and $\mathcal{Z}_{y}$ resembles
the path integral of a particle moving in 1-dimensional harmonic potential.
This is more obvious if we apply the following substitutions:

\begin{equation}
s\leftrightarrow i\xi,\quad\beta\leftrightarrow1/\hbar,\quad h_{11}\leftrightarrow m,\quad F\leftrightarrow m\omega^{2},\label{eq:substitution}
\end{equation}

Then $\mathcal{Z}_{t}$ takes exactly the form of a path integral:

\begin{equation}
\mathcal{Z}_{t}=\int_{C}\left[Dt\right]\exp\left[\frac{i}{\hbar}\int_{0}^{-iL_{0}}\left(\frac{m}{2}\left(\frac{dt}{d\xi}\right)^{2}-\frac{m\omega^{2}}{2}t^{2}\right)d\xi\right].\label{eq:pathintegral}
\end{equation}

The path integral above is one of the few examples where an analytical
solution can be derived. We just use the known result here and obtain
\begin{equation}
{\cal Z}_{t}=\sqrt{\frac{m\omega}{2\pi i\hbar\sin\left(\omega\left(\xi_{f}-\xi_{i}\right)\right)}}\exp\left[\frac{i}{2\hbar}m\omega\frac{\left(t_{i}^{2}+t_{f}^{2}\right)\cos\left(\omega\left(\xi_{f}-\xi_{i}\right)\right)-2t_{i}t_{f}}{\sin\left(\omega\left(\xi_{f}-\xi_{i}\right)\right)}\right].\label{eq:pathintegral-result}
\end{equation}

The initial and final values $t_{i}$ and $t_{f}$ in the equation
above corresponds to the boundary conditions at the two ends of the
chain. For different situations, different boundary conditions can
be used, but the overall complexity of the problem will not be increased. Here, as mentioned in the main text, we choose
the boundary conditions where the tangent vectors at the two ends
are parallel to the force direction, i.e. $t_{i}=t_{f}=0$. Using
Eq. (\ref{eq:substitution}) and also noting that $\xi_{i}=0$
and $\xi_{f}=-iL_{0}$, we have
\begin{equation}
\mathcal{Z}_{t}=(h_{11}F)^{1/4}\left(\frac{\beta}{2\pi\sinh\sqrt{FL_{0}^{2}/h_{11}}}\right)^{1/2}
\end{equation}
which gives the complete partition function
\begin{equation}
\mathcal{Z}=e^{\beta FL_{0}}\mathcal{Z}_{t}^{2}=e^{\beta FL_{0}}(h_{11}F)^{1/2}\left(\frac{\beta}{2\pi\sinh\sqrt{FL_{0}^{2}/h_{11}}}\right).\label{eq:partitionanalyticAPP}
\end{equation}
It should be noted that the above calculations can be easily generalized to any space dimensions $d$.

\vspace*{-3pt}
%


\end{document}